# Une approche pour mieux appréhender l'altérité en SIC

**Pierre-Michel RICCIO**
LGI2P – Ecole des mines d'Alès
Pierre-Michel.Riccio@mines-ales.fr

**RÉSUMÉ**
Dans cet article nous proposons une démarche originale qui a pour ambition : de faciliter la construction d'équipes en s'appuyant sur des notions de compétences, de motivation, ou de potentiel ; d'identifier l'effort de mise à niveau des connaissances nécessaire pour intégrer de façon pragmatique une équipe ; ou de créer des dispositifs mieux adaptés pour une population cible. L'ensemble forme une boite à outils qui nous semble adaptée pour faciliter une meilleure prise en compte de l'altérité.

**Mots-Clés :**
Sic, dispositif, méthodologie, prototype, classe, agrégation multicritère

**ABSTRACT**
In this paper we propose a novel approach that aims: to facilitate the building of teams relying on notions of skills, of motivation or of potential; identify the requested upgrade effort to improve his own knowledge and join a team; or create more suitable devices for a target population. The whole forms a toolbox that seems appropriate to facilitate better recognition of otherness.

**INTRODUCTION**

A l'heure du développement des technologies numériques, l'objectif de cet article est de prendre du recul sur notre façon de construire les dispositifs de communication afin d'identifier les pistes qui permettraient d'aménager concepts méthodes et outils pour mieux prendre en compte l'altérité.

Dans un contexte de généralisation des technologies numériques et de renforcement des échanges entre individus, que ceux-ci soient résidants ou récemment installés, nous pouvons constater un foisonnement des interactions entre initiés et une mise à l'écart de fait d'un grand nombre de personnes. Dans leurs activités professionnelles ou personnelles de nombreux acteurs, pour différentes raisons, se retrouvent exclus de la vie numérique.

L'objet de cet article n'est pas de détailler les raisons de l'exclusion, mais de porter un regard sur la façon dont nous abordons en sciences humaines et sociales (SHS) et plus particulièrement en sciences de l'information et de la communication (SIC) la construction de dispositifs de communication, puis de repérer comment mieux appréhender l'altérité.

**1 – LA PROBLEMATIQUE**

Le travail réalisé ces dernières années en SHS et en SIC a permis d'améliorer dans de larges proportions la qualité des dispositifs numériques de communication (Riccio, 2010).

Le fait de mieux comprendre les enjeux et normes portés par les acteurs en situation à l'aide par exemple de la théorie sémio-contextuelle des communications (Mucchielli, 2000) ou de mieux appréhender les luttes de pouvoir et de positionnement des différents acteurs à l'aide par exemple de la théorie acteur-réseau (Akrich Callon et Latour, 2006) facilite désormais la construction de dispositifs de qualité. Ces derniers sont aujourd'hui beaucoup plus satisfaisants : pour les utilisateurs, comme pour les équipes de conception.

A noter que nous nous appuyons ici sur la notion de dispositif proposée par Michel Foucault : « un ensemble résolument hétérogène, comportant des discours, des institutions, des aménagements architecturaux, des décisions réglementaires, des lois, des mesures administratives, des énoncés scientifiques, des propositions philosophiques, morales, philanthropiques, bref : du dit, aussi bien que du non-dit » (Foucault, 1994).

Mais, les approches utilisées permettent-elles de prendre en compte l'altérité, entendue comme une appréhension de ce qui est autre, de ce qui est différent ?

La réponse est complexe.

Nous avons travaillé par exemple sur la conception de dispositifs destinés à des populations fragiles : travail centré sur la mise au point d'un dispositif de rééducation fonctionnelle pour les déficients visuels (Vasquez, 2000). Le dispositif créé est un succès, il est aujourd'hui utilisé en Europe comme en Amérique du nord.

Mais en prenant du recul, la démarche que avons mise en œuvre - une combinaison d'approches sémio-contextuelle et acteur-réseau basée sur une validation par saturation qualitative et triangulation - permet avant tout de traiter des groupes assez homogènes. Elle laisse peu de place pour appréhender une distribution de l'altérité.

En fait, ces théories – sans le revendiquer de façon explicite – sont construites sur la base de la notion de classe. Rappelons que la classe peut être définie comme : « un ensemble d'individus possédant la même structure et le même comportement » (Stefik & Bobrow, 1986). L'identification en situation des attributs (des propriétés) et des individus permet de construire des groupes d'acteurs. Un individu dès qu'il possède l'ensemble des attributs qui caractérisent une classe est alors rattaché à cette classe, par exemple : clients, commerçants, concurrents …

Ces théories permettent aussi d'identifier les individus qui possèdent un poids particulier en regard de l'une des dimensions (attributs ou propriétés) propre à la grille de lecture utilisée comme : les enjeux, les normes, le pouvoir, ou encore la position. Mais il est beaucoup plus difficile de prendre en compte les individus qui n'appartiennent pas à une classe, sans pour autant que leurs propriétés intrinsèques attire l'attention de l'analyste.

Aussi, que fait-on dans ce cas des individus – comme l'ornithorynque d'Umberto Eco (Eco, 1997) – qui appartiennent à plusieurs classes et *a fortiori* des individus qui n'appartiennent à aucune des classes ?

Comment à l'aide de ces théories peut-on appréhender l'altérité ?

**2 – LE CONTEXTE**

Pour permettre au lecteur de comprendre plus facilement notre démarche, nous allons maintenant présenter de façon synthétique les trois « théories » sur lesquelles nous nous appuyons.

**2.1 - La théorie sémio-contextuelle**

L'objet de la théorie sémio-contextuelle des communications (Mucchielli, 2000) est de faire émerger, dans une approche systémique et constructiviste (Morin, 1994), le sens qui accompagne toute communication généralisée, expression d'intentionnalités explicites ou latentes dans une situation d'échange par et pour des acteurs.

L'annotation sémio-contextuelle (Riccio, 2003) consiste à repérer dans le récit – issu de la mise au net des éléments d'informations collectés – les processus de communication et l'impact de ces processus sur les contextes de la situation (spatial, physique, temporel, position, normes, relations, enjeux). Ces annotations, à situer au plus près du texte, facilitent l'émergence du sens et l'analyse qui va permettre la compréhension générale de la situation.

*Le contexte spatial*

Le lieu de réunion, l'aménagement de la salle, la position des individus autour de la table, la distance entre les personnes, les moyens techniques, etc. sont des composantes qui évoquées ou manipulées par les acteurs d'une situation influencent le sens des communications.

*Le contexte physique et sensoriel*

Des impacts sensoriels multiples (visuels, sonores, olfactifs, tactiles) en combinaison avec un contexte spatial organisé et des acteurs prédisposés, peuvent transformer les modalités de perception et la signification des communications.

*Le contexte temporel*

Toute communication généralisée s'inscrit dans un contexte historique (elle vient se positionner dans le cadre de ce qui a déjà été communiqué) et dynamique (elle se caractérise par un rythme, régulier ou en rupture, et par des appels au temps, explicites ou implicites).

*Le contexte des positions respectives des acteurs*

Chaque individu propose à travers son langage (tutoiement / vouvoiement) mais aussi son paralangage (habillement) sa vision du positionnement des acteurs de la situation. L'échange est une lutte permanente dans laquelle tout individu cherche à valoriser sa position.

*Le contexte culturel de référence aux normes*

Que ce soit dans la vie professionnelle ou dans la vie personnelle, les normes (culturelles et sociales) et les règles (règlements, pratiques et usages) forment un « déjà là » et définissent un sens *a priori* partagé par un groupe d'individus.

*Le contexte relationnel immédiat*

Chaque individu utilise le langage et le paralangage pour séduire et influencer ses interlocuteurs. Pour faire émerger le sens, il est nécessaire d'identifier les composantes valorisantes et dévalorisantes mises en œuvre dans l'échange.

*Le contexte expressif de l'identité des acteurs*

Tout individu est doté d'un système de pertinence qui en fonction de ses préoccupations forme sa vision du monde, c'est-à-dire une perception sélective des phénomènes de la vie (Schutz, 1994). Il n'est pas possible de saisir le sens d'une communication sans comprendre la motivation des acteurs dans la situation, leurs enjeux.

**2.2. - La théorie de l'acteur-réseau**

La théorie de l'acteur réseau est une théorie originale développée sous l'impulsion du Centre de sociologie de l'innovation (CSI) de l'Ecole des mines de Paris afin d'analyser les différentes façons dont société et sciences se mélangent (Akrich Callon et Latour, 2006). Elle propose de nombreux outils pour la gestion de l'innovation et le suivi des transformations techniques.

L'originalité de la théorie de l'acteur réseau, qui s'appuie sur le concept d'acteur, est de considérer qu'il n'y a pas dans une situation de micro-acteurs et de macro-acteurs, mais un ensemble d'acteurs isomorphes (Callon, 2006). L'isomorphie ne signifie pas que tous les acteurs ont la même taille, mais que la taille ne peut pas être décidée a priori car elle est le résultat de processus de négociation achevés ou en-cours.

La pratique générale consiste à utiliser des grilles d'analyse différentes pour étudier les micro-acteurs et les macro-acteurs. Mais, un cadrage trop étroit peut conduire à des interprétations erronées, des contresens. Un cadrage trop large ne permet pas d'appréhender l'ensemble des phénomènes. Aussi, la théorie de l'acteur réseau propose d'utiliser la même grille d'analyse pour appréhender l'ensemble des acteurs impliqués dans la situation.

L'étude est « simplifiée » grâce à l'utilisation d'un mécanisme appelé : « boîte noire ». Dans cette approche, le macro-acteur est un micro-acteur assis sur des boîtes noires. Les boîtes noires sont composées de raisonnements, d'habitudes, de forces, d'objets, résultats d'une négociation et stabilisés.

Le travail consiste alors à analyser les opérations par lesquelles un acteur crée des asymétries plus ou moins durables. Celles-ci peuvent être expliquées grâce au concept de traduction, développé par Michel Serres (Serres, 1974) et repris par Michel Callon (Callon, 1975). Par traduction est entendu : « l'ensemble des négociations, des intrigues, des actes de persuasion, des calculs, des violences grâce à quoi un acteur ou une force se permet ou se fait attribuer l'autorité de parler ou d'agir au nom d'un autre acteur ou d'une autre force » (Callon et Latour, 2006).

Le fait de centrer l'étude sur l'identification des « négociations » dans le cours d'action et d'en inférer le positionnement des acteurs est une voie intéressante pour comprendre plus facilement et plus rapidement les composantes essentielles de la situation.

### 2.3 - La théorie des prototypes

Des Grecs de l'antiquité et plus particulièrement d'Aristote, nous avons hérité d'une « tradition » de classification des objets du monde. Les objets sont organisés en groupes ou classes, chaque classe étant caractérisée par un ensemble de propriétés.

Cette « tradition » culturelle, élément essentiel de nos apprentissages, est à la base de notre perception du monde : animaux, éléments naturels, ou objets de la vie courante. Si un objet (par exemple ma voiture) appartient à une classe (voiture de sport) il en possède alors sans exception toutes les propriétés (axiome du tiers exclu).

S'il faut reconnaître que la classification aristotélicienne a facilité pendant plusieurs siècles le partage et la redistribution des savoirs, le nombre croissant d'exceptions – de l'ornithorynque aux connaissances partagées par un collectif (Eco, 1997) – dans un monde complexe en réseaux, a conduit ces dernières années de nombreux spécialistes à s'interroger sur le mécanisme de classification et son impact dans différents domaines d'application. Quelle peut-être la valeur d'un modèle sans une vision diachronique du système (Le Moigne, 1990) ?

De père anglais et de mère russe, Eleanor Rosch a grandi aux Etats-Unis. Diplômée de Harvard à la fin des années 60, elle devient ensuite professeur au département de Psychologie à Université de Californie à Berkeley. Au terme de sa formation, elle décide de partir étudier avec son mari anthropologue la population des Danis en Nouvelle Guinée. Elle effectuera deux séjours en immersion. Son travail a porté d'une part sur les catégories de couleurs et de formes chez ces indigènes, d'autre part sur l'éducation des enfants qui est, selon elle : « un ensemble d'événements mémorisés d'interactions entre mères et enfants ».

Il existait jusqu'au milieu du XXème siècle, une vision dominante de la classification des objets du monde : l'esprit « occidental » était capable d'abstraire le savoir des idiosyncrasies de l'expérience individuelle quotidienne et, ce faisant, utilisait les lois aristotéliciennes de la logique.

Appliqué à la classification, cela signifiait que pour connaître une catégorie (ou un ensemble de classes) il fallait avoir des critères abstraits, précis, nécessaires et suffisants pour définir les éléments qu'elle incluait (i.e. pour énoncer ses propriétés) (Rosch et Lloyd, 1978).

Passionnée par la problématique langage / perception, Eleanor Rosch, qui travaillait depuis plusieurs années avec Roger Brown (psycholinguiste spécialiste du « spectre des couleurs ») à Harvard, décide d'observer *in situ* une population indigène de l'âge de pierre, les Danis de Nouvelle Guinée, dont la particularité est de n'utiliser que deux noms pour désigner l'ensemble des couleurs : *mola* pour les nuances brillantes et chaudes, et *mili* pour les nuances froides et sombres.

Dans une première expérimentation, Eleanor Rosch présente aux Danis 40 pièces de couleur (4 niveaux de brillance et 10 niveaux de teinte), et leur demande de nommer les pièces. Il faut préciser qu'à cette époque de nombreux chercheurs considéraient que la séparation entre couleurs était arbitraire, aboutissement d'un processus culturel traduit sous une forme linguistique. A chaque extrémité du spectre des couleurs les Danis étaient d'accord entre eux, et, même si le consensus n'était pas complet concernant les couleurs intermédiaires, les premiers essais confirmaient une évidence : les Danis possèdent une culture différente de la nôtre.

Dans la deuxième expérimentation, Eleanor Rosch demande aux Danis de reconnaître une pièce de couleur. Elle présente à chaque individu une pièce, lui demande de patienter dans l'obscurité, puis de retrouver cette pièce parmi l'ensemble des pièces disponibles. Dans le contexte général, les résultats sont particulièrement étonnants : les Danis reconnaissent les couleurs d'une façon très semblable à celle des « occidentaux ». Roger Brown précise : « L'ironie fascinante de l'histoire est que cette recherche a commencé dans un esprit de fort relativisme et de déterminisme linguistique, et qu'elle arrive à la conclusion de l'universalisme culturel de l'insignifiance linguistique. » (Brown, 1975).

Eleanor Rosch réalisera ensuite de nombreuses expérimentations, pour arriver à la conclusion que les catégories sont construites autour d'un élément central qu'elle appellera : prototype. La principale caractéristique d'un prototype est de partager de nombreuses propriétés avec certains objets du monde (qui forment une catégorie) et peu avec les autres objets (qui de fait appartiennent à d'autres catégories).

Nous ne sommes plus dans le cadre où, pour appartenir à une classe, l'objet doit posséder l'ensemble exhaustif des attributs de cette classe. Nous sommes plutôt dans une logique « floue » ou la composition de la catégorie est déterminée par une relative proximité à un objet émergeant : le prototype.

## 3 – LES PISTES POUR MIEUX PRENDRE EN COMPTE L'ALTERITE

En préambule à l'esquisse des différentes pistes qui vont nous permettre de mieux prendre en compte l'altérité, il nous semble important de préciser que notre démarche n'est pas de construire une ou des classes dans lesquelles nous allons grouper ceux qui sont différents. Mais bien d'identifier la capacité des individus, dans un ensemble, à conduire une tâche pour laquelle ils ne possèdent pas toutes les qualités *a priori.*

**3.1 – Première piste : mise en œuvre de la théorie des prototypes**

La première piste – sans modifier le fondement des approches sémio-contextuelle et acteur-réseau qui ont prouvé leur efficacité – est de trouver une nouvelle façon de construire des classes d'acteurs, mieux orientée sur la possibilité de conduire ou de participer à une tâche au moment de l'action. Ceci permettra d'éviter que des individus qui ne possèdent pas toutes les qualités requises soient écartés dans le processus.

Pour ce faire, nous proposons de convoquer la théorie des prototypes imaginée par Eléanor Rosch afin de construire les catégories autour d'un élément central appelé prototype (Rosch, 1975).

La principale caractéristique d'un prototype est de partager des propriétés avec certains objets du monde (qui forment une catégorie ou classe) et peu avec les autres objets (qui forment d'autres classes). Nous ne sommes plus dans le cadre où pour appartenir à une classe, l'objet doit posséder l'ensemble exhaustif des attributs de cette classe. Nous sommes dans une logique « floue » où la composition de la classe est déterminée par une proximité relative des d'objets ou individus à un prototype qui de fait constitue le centre de gravité de cette classe.

Il est donc nécessaire d'identifier quelles sont les qualités nécessaires pour conduire ou participer à une tâche. Selon l'origine et les connaissances de l'analyste, son système de pertinence (Schutz, 1994), le terme qualité peut être remplacé par d'autres termes : dimensions, critères ou attributs.

Il est ensuite nécessaire d'évaluer de façon quantitative ou qualitative pour chaque individu et pour chaque qualité un niveau. L'échelle est ici importante, nous reviendrons sur cette question un peu plus tard.

Bien entendu, le prototype qui matérialise le centre de gravité de la classe est un individu réel ou virtuel qui possède les meilleurs résultats possibles sur l'ensemble des qualités propres à cette classe.

Chaque individu – dont le profil forme un vecteur – peut être alors positionné à une certaine distance du prototype. Cette cartographie évolue en permanence, à travers la valorisation des qualités des individus qui vont elles-mêmes évoluer dans le temps.

Ce n'est qu'au moment de l'action, voire même en projetant les profils des individus dans le temps de l'action, que les limites des différentes classes seront alors définies.

**3.2 – Deuxième piste : appui sur une théorie de l'utilisation**

La deuxième piste est qu'il convient de déterminer quelle combinaison de qualités (valorisation des dimensions, critères ou attributs) est nécessaire pour définir une classe, et au-delà chacune des classes.

La proximité d'un individu par rapport à un ou plusieurs prototypes (et de fait son appartenance à une classe donnée) peut alors être évaluée par comparaison entre son profil – résultat de l'agrégation de l'ensemble de ses « qualités » – et le profil du ou des prototypes les plus proches.

Pour ce faire nous proposons de nous appuyer sur une théorie de l'utilisation qui postule que pour un dispositif l'utilisation résulte de la combinaison de pratiques, identifiées comme l'ensemble des compétences et motivations portées par l'individu en situation, et d'usages, identifiés comme l'ensemble des fonctions mises à disposition des utilisateurs.

En ce qui concerne les usages : le dispositif, dans le sens de Michel Foucault, est conçu pour proposer un ensemble de fonctions aux utilisateurs. La machine à café ne sert *a priori* qu'à faire du café, mais un robot ménager proposera par exemple une assez large palette de fonctions. De nombreuses études, comme celles de Victor Scardigli (Scardigli, 1992) ont montré que la relation entre fonctions et utilisateur peut être assez complexe. L'utilisateur peut refuser d'utiliser certaines fonctions, ne pas être capable de les utiliser, ou encore les détourner pour faire autre chose. Dans le même esprit les concepteurs peuvent oublier de mettre à disposition certaines fonctions, choisir de ne pas les déployer, ou encore les mettre à disposition de façon inutile car par exemple masquées ou non documentées. La question de l'appropriation d'un dispositif est donc de fait des usages est loin d'être simple.

En ce qui concerne les pratiques : comme le précise Alfred Schutz chaque individu est doté d'un système de pertinence qui à travers l'ensemble de ses apprentissages forme sa vision du monde et de fait sa capacité à faire ou ne pas faire. Aussi, nous proposons le parti pris suivant qui est de considérer que les pratiques sont constituées pour chaque individu des compétences acquises et de sa motivation à faire. Au moment où l'Homme est amené durant sa carrière professionnelle à exercer des activités qui peuvent être très différentes, cela pose la question du potentiel de l'individu et de l'ajustement de celui-ci à travers par exemple des actions destinées à compléter sa formation ou à renforcer sa motivation.

L'utilisation optimale devient alors une question de mise en adéquation des usages et pratiques dans une approche diachronique. Il ne s'agit plus uniquement d'évaluer un ensemble à un moment donné qui est celui du départ de l'action, mais de projeter l'ensemble des dimensions dans le temps pour maximiser le résultat final : la réussite de l'action.

Cette démarche qui peut sembler complexe présente toutefois un avantage indéniable : la possibilité de mieux intégrer dans un processus dynamique ceux qui sont au départ différents mais possèdent un certain potentiel. En jouant sur la formation, sur la motivation, mais aussi sur la nature des fonctions opérationnelles du dispositif à déployer, la démarche contribue à une meilleure prise en compte de l'altérité.

La construction des classes peut alors s'effectuer à travers un repérage d'un ensemble de qualités liées à la compétence, la motivation et la capacité à utiliser certaines fonctions des dispositifs.

### 3.3 – Troisième piste : assemblage par agrégation multicritère

La troisième piste réside dans la façon d'assembler les différents éléments.

En effet pour composer les classes, il est nécessaire d'évaluer la distance entre chaque individu et les différents prototypes afin de déterminer quels sont les individus qui vont être rattachés à chacune des classes. Aussi, il est nécessaire d'imaginer une fonction d'agrégation.

Chaque qualité (dimension ou critère) peut être pondérée, et la fonction naturelle d'agrégation généralement utilisée est la moyenne pondérée. Mais cette fonction ne permet pas toujours de traduire un objectif complexe.

Par exemple, celle-ci est peu adaptée pour traduire le besoin suivant : « pour compléter mon équipe je souhaite retenir une personne qui possède de bonnes connaissances en mathématiques et un assez bon niveau de connaissances en français ou bien qui peut acquérir ces connaissances assez rapidement ».

Il existe d'autres fonctions d'agrégation comme par exemple l'intégrale de Choquet 2-additive. Cette dernière est un peu plus complexe, mais elle permet de traduire de façon beaucoup plus fine l'expression d'un besoin en s'appuyant sur une évaluation quantitative et / ou qualitative des critères mesurés à un moment donné (celui de l'action) tout en étant capable de prendre en compte dans l'agrégation le potentiel de progression identifié en amont.

Sans trop entrer dans les fonctions mathématiques, ce que nous ferons prochainement dans une version étendue de cet article, l'idée est valoriser une cohérence d'ensemble au détriment des particularités spécifiques.

Voici un exemple simplifié. Pour compléter ou composer notre équipe nous avons le choix entre trois individus (évaluation basée sur les capacités et le potentiel) : le premier est très bon en math et moyen en français, le deuxième moyen en math et très bon en français, le troisième bon en math et en français. Si nous attribuons aux trois individus les scores suivants (20, 10), (10, 20) et (15, 15) nous voyons bien que la moyenne pondérée va nous donner le même résultat. Or, nous percevons bien que le troisième individu serait plus intéressant à recruter. Il est donc nécessaire de trouver une nouvelle fonction d'agrégation qui va renforcer la cohérence de l'assemblage. De façon schématique l'intégrale de Choquet 2-additive permet de retrancher une partie de la différence entre les valeurs extrêmes à la moyenne. De fait, le troisième profil plus cohérent va naturellement émerger.

## 4 – DISCUSSION

Nous venons de présenter trois pistes qui combinées permettent à notre avis de faciliter la prise en compte de l'altérité.

La première piste permet de construire des classes souples, qui ne sont plus fondées sur le fait de posséder ou non une qualité, mais bien sur une combinaison de l'ensemble des qualités pour conduire une tâche ou rejoindre une équipe.

La seconde piste donne un guide pour identifier les qualités (dimensions ou critères) nécessaires pour évaluer l'aptitude d'un individu à rejoindre une classe à un moment donné. Ces qualités qui sont fondées sur les compétences, les motivations, la capacité à utiliser les fonctions d'un dispositif, sont centrées sur la recherche du meilleur assemblage possible pour réaliser une tâche avec succès.

Enfin, la troisième piste propose une façon originale d'assembler les individus pour composer une équipe.

La question est maintenant de savoir quel peut l'impact de cette façon de procéder sur les approches classiques en Sciences de l'Information et de la Communication afin d'élaborer des dispositifs plus ouverts à l'altérité.

Dès lors que nous ne sommes pas dans le cas d'une micro-situation, c'est-à-dire d'une étude portant sur un nombre très limité de faits et d'acteurs, il est généralement nécessaire de conduire une enquête assez large pour disposer des éléments indispensables à la compréhension de la situation.

Les techniques pour conduire ce type d'enquête sont depuis longtemps maîtrisées : observation, observation participante, entretiens individuels non directifs centrés, entretiens qualitatifs de groupes, étude des traces …

Mais il est d'expérience assez difficile de tout observer, d'interviewer l'ensemble des acteurs. Cela conduit assez rapidement l'analyste à composer des groupes, des classes, avec toujours un doute sur la couverture générale : est-ce que j'ai bien rassemblé l'ensemble des points de vue pour une situation donnée ?

Nous savons bien que la qualité de l'analyse est largement dépendante de la qualité des données et informations collectées. Or, le mécanisme cité ci-dessus laisse peu de place à l'altérité.

Sans constituer une baguette magique, l'approche que nous proposons peut à notre avis grandement améliorer le résultat final.

Et cela peut se faire de différentes façons :
- en facilitant la composition des groupes à interviewer,
- en facilitant le repérage des « forces » en présence,
- en identifiant en amont des minorités pertinentes (dont l'avis peut être important),
- …

De fait il ne s'agit plus de chercher à repérer dans la situation des groupes d'acteurs, mais bien des prototypes autour desquels l'ensemble des acteurs vont pouvoir être positionnés, et si nécessaire repositionnés.

## CONCLUSION

Nous proposons une approche nouvelle pour la composition des groupes d'acteurs ou la création de dispositifs.

Les groupes d'acteurs peuvent être créés par rapport à un objectif pragmatique en s'appuyant sur des critères comme la compétence, la motivation et la capacité à utiliser un dispositif. Dégagée de critères comme le physique, ou l'origine, cette démarche facilite la prise en compte de l'altérité.

De plus, dans un processus inverse elle permet d'appréhender quels sont les points à améliorer pour intégrer plus facilement un groupe ou une équipe : la mise à niveau ou la formation des individus ; ou encore la façon de les motiver à entreprendre.

En partant cette fois du centre de gravité (ou prototype) il est aussi possible d'imaginer quelle sont les fonctions d'un dispositif à construire et déployer pour – dans un processus inverse – adapter celui-ci à la population cible.

Enfin, que ce soit dans la collecte des informations ou dans leur analyse, les concepts méthodes et outils des SIC peuvent être aménagés pour intégrer cette démarche dans un ensemble méthodologique.

## BIBLIOGRAPHIE